\def\astroph{1}      % astro-ph or not
\def\epsgz{0}        % gzip'ed .eps files (dvipdf may not work) or not

\ifnum\astroph=0                        % ApJL submission
\documentclass[10pt,preprint2]{aastex} % Double column, single spaced

\else                                   % ApJL emulation, for astro-ph 
\documentclass{emulateapj} 
\usepackage{apjfonts} 
\fi

\usepackage{epsfig,graphicx,natbib}

\newcommand{\fig}[1]{Fig.\ \ref{#1}}

\slugcomment{Accepted by Astrophysical Journal Letters.}

\begin{document}

\shorttitle{Stochastically Induced GRB Wakefields} 
\shortauthors{J. Trier Frederiksen}

\title{Stochastically Induced Gamma-Ray Burst Wakefield Processes}
\author{Jacob Trier Frederiksen} 
\affil{Stockholm Observatory, AlbaNova University Center, SE-106\,91 Stockholm, Sweden} 
\email{jacob@astro.su.se / trier@astro.ku.dk}

%---------------------------------------------------------------------
\begin{abstract} We present a numerical study of Gamma-Ray Burst -
Circumburst Medium  interaction and plasma preconditioning via Compton
scattering. The simulation tool employed here -- the
\textsc{PhotonPlasma} code -- is a unique hybrid model; it combines a
highly parallelized (Vlasov) particle-in-cell approach with continuous
weighting of particles and a sub-Debye Monte-Carlo binary particle
interaction framework. Our first results from 3D simulations with this
new simulation tool suggest that magnetic fields and plasma density
filaments are created in the wakefield of prompt gamma-ray bursts, and
that the photon flux density gradient has a significant impact on
particle acceleration in the burst head and wakefield. We discuss some
possible implications of the circumburst medium preconditioning for
the trailing afterglow, and also discuss which additional processes
will be needed to improve future studies within this unique and
powerful simulation framework. 
\end{abstract}


%---------------------------------------------------------------------
\section{Introduction} The microphysics of Gamma-Ray Burst (GRB)
afterglow shocks propagating through a circumburst medium (CBM) has
been theoretically predicted by \citep{bib:Medvedev1999} and later 
extensively studied using particle-in-cell (PIC) models in the
past few years \citep[see e.g.][]{bib:Spitkovsky2008,bib:Hededal2004,
bib:Frederiksen2004,bib:Silva2003,bib:Gruzinov2001a}. 
In contrast, the initial interaction between the
GRB photons and the CBM plasma is less well studied numerically,
although progress has been made for coherent wakefield processes in
lower-dimensional and pair plasmas
\citep[e.g.][]{bib:Hoshino2008,bib:Liang2007}. Whereas the burst-CBM
interaction has received strong attention in
theoretical~\citep{bib:Beloborodov2002,bib:Madau2000} and 
observational~\citep{bib:Connaughton2002} works, the detailed 
microphysical plasma dynamics in the first pulse-plasma 
encounter lacks sufficient treatment with respect to the high degree 
of forcing anisotropy in the problem of photon scattering off the 
relatively cold tenuous CBM.

Wakefield interactions between high energy photon pulses and a plasma
have been extensively studied in the context of \emph{coherent}
wakefield acceleration in laser-plasma systems, both experimentally 
\citep[e.g.][]{bib:Phuoc2005,bib:Kostyukov2003}, and numerically in 
the same context~\citep[e.g.][]{bib:Phuoc2005,bib:Pukhov1999}. 
In the astrophysical context of particle acceleration in GRBs, 
wakefield acceleration has been studied in the coherent regime where 
the forcing of the plasma is done by the electromagnetic field 
\citep{bib:Hoshino2008,bib:Liang2007}. Incoherent -- or 
\emph{stochastic} -- wakefield processes have to our knowledge so far 
not been studied numerically, and has been considered theoretically 
only in a few cases 
\citep[e.g.][]{bib:Barbiellini2006,bib:Barbiellini2004}.

Our motivation for developing a new and improved Particle-in-cell-
Monte-Carlo (PIC-MC) code framework for investigating the detailed
Compton interaction of an ultra-intense GRB and a quiescent cold CBM
is two-fold:

First, due to the high-energy spectrum of the prompt GRB (for BATSE of
order $E_\gamma \sim m_e c^2$), the photon wavelength is much shorter
than any characteristic length scale in the CBM plasma,
$\lambda_\gamma \ll \delta_e$ and $\lambda_\gamma \ll \lambda_D$, where 
$\delta_e$ and $\lambda_D$ are the electron skin-depth and Debye-length, 
respectively. We therefore expect a binary approximation to work well for
prompt burst photons interacting with the plasma constituent particles
through Compton scattering. In fact, photons in a large range down
towards the electron plasma frequency
$\omega_\gamma \sim \omega_{pe} = \lambda_D^{-1} v_{th,e}$ (where
$v_{th,e}$ is the electron thermal speed) could be treated as point
particles that interact in a binary way with the plasma.

Secondly, the extreme intensity of the burst makes the GRB front
opaque to the plasma electrons. Assume a CBM with
$n_e \approx 1~\textnormal{cm}^{-3}$, and further assume a typical
BATSE GRB with an estimated~\citep[from][eqn.2]{bib:Barbiellini2006}
photon flux energy density of
$10^{35}~\textnormal{eV}~\textnormal{cm}^{-2}~\textnormal{s}^{-1}$ at
distance $R_0 \sim 10^{16}~\textnormal{cm}$ from the progenitor. 
Then, Compton scattering off the CBM plasma will occur many times 
every second per electron during the prompt phase; most
photons survive traversing the plasma without being significantly
affected. The plasma is forced predominantly by binary particle-
particle encounters, everywhere locally.

These considerations motivated us to develop and employ a new and
unique tool in order to gain access to the sub-Debye regime. Adopting
the "random phase approximation"~\citep{bib:Pines1952} we may split,
in a natural way, plasma processes into two different numerical
schemes depending on the characteristic scale of the dynamics, here
denoted $\textbf{k}_C$: \begin{description} \item[
$\lambda_D \ll \left|{\bf k}_C\right|^{-1}$:] for forcing and
disturbances scale larger than the Debye length the Vlasov approach is
applied and the PIC code framework is utilized. \item[
$\lambda_D \gg \left|{\bf k}_C\right|^{-1}$:] in the opposite case
dynamics is particle-particle interaction dominated, and we solve the
scattering problem statistically through MC radiative transfer for the
plasma constituents. \end{description}

\noindent In this way we obtain a substantially improved description
of the plasma for both the Vlasov and detailed balance cases, and
capture in a natural way the details of wave-particle interaction and
scattering. The \textsc{PhotonPlasma} code is parallelized with MPI
and is highly scalable. Any plasma constituent species, forcing
and scattering process is possible to incorporate with relative ease;
for further details see \citep{bib:Haugboelle2005,bib:Frederiksen2008}.

In the remainder of this Letter we present, in Section 2, some central
details on our numerical approach, and account for the simulation
setup as well as physical and numerical scaling. Results and
discussion are given in Section 3, where we also relate the wakefield
effects to other results concerning coherent -- rather than
stochastically induced -- wakefield processes. Conclusions are
given in Section 4.

%---------------------------------------------------------------------
\section{The Stochastic Wakefield Simulation} Here we describe the 3D
runs from our first wakefield experiments, full details of this as
well as other setups, tests and more results can be found in
\cite{bib:Frederiksen2008}.

\begin{figure}[!ht] 
\centering 
\ifnum\epsgz=1 
 \epsfig{figure=f1.eps.gz,width=\linewidth} 
\else 
 \epsfig{figure=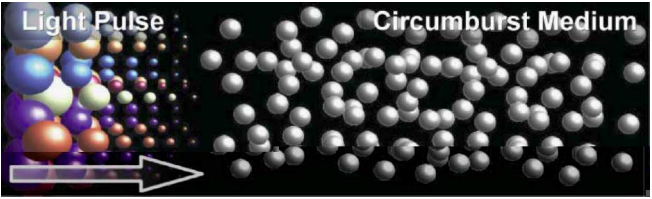,width=\linewidth} 
\fi 
\caption{Illustration of our computational setup; GRB photons (left) move 
through the ambient CBM plasma. As the intensity of the light 
pulse increases, the \textit{weights} (size) of the photons vary, 
rather than the \textit{number}. Color designates that photons carry 
different energies (through $\nu$) as well.} 
\label{fig:lightpulse} 
\end{figure}

A synthetic thermal GRB is delivered to a quiescent CBM plasma by
adding photons (particles) to the computational domain on a volume
boundary, as illustrated in figure~\fig{fig:lightpulse}, according to
a prescribed light curve and spectrum.

The light curve is inspired by BATSE data fitted to a FRED function~
\citep{bib:Kocevski2003,bib:Ryde2002} with parameters chosen for BATSE
trigger 3891, $r=1.26$ and $d=2.67$. The spectral window
coincides with the BATSE window,
$E_\gamma \in [33.6\textnormal{keV},3.36\textnormal{MeV}]$ -- after
correction for an estimated redshift of $z \sim 1.68$. The spectrum 
is assumed to be Planckian: 
$N_{\gamma}(\nu,t) = (h\nu)^{-1}F(\nu,t) = (h\nu)^{-1}B_\nu(T(t),t)$, for
the photon spectral radiance. We use the Stefan-Boltzmann law to model
the temperature, $T(t) \propto F(t)^{1/4}$. Our model 
FRED has duration $\tau_{GRB} = 200$, corresponding to about 
1\% of $T_{50}$ for trigger 3891.

We evolve our synthetic GRB by adding a \textit{constant} number of photons per
cell carrying \textit{continuous} weights prescribed by the 
burst functions. Photon energies are sampled from MC
integration of the (time-dependent) Planckian. The photon
spectral radiance is then 
$$n_\gamma(\nu,t) \propto \frac{F(\nu,t)}{\nu} \propto 
\sum^{N_{\gamma}(\nu,t)}_{i=1} w_i(\nu,t) ~ \Rightarrow ~ n_\gamma(\nu,t) 
\propto \sum^{N_{\gamma}(\nu)}_{i=1} w_i(t)~,$$ 
using our assumptions about the thermal GRB. 
The last sum, split into assigning time-variable weights and 
frequency variability to individual particles, gives a high degree
of flexibility compared to conventional PIC codes.

The CBM plasma is assumed to be a uniform density medium, consisting 
of a hydrogenic, fully ionized, moderate temperature plasma with 
$v_{th,e} \approx 0.1c$. We have adopted conventional PIC scaling, 
setting all relevant natural constants equal to unity, except the 
ion-to-electron mass ratio, which is  $m_i/m_e = 256$. The initial 
number of particles is 40/cell/species, and the grid resolution 
100$\times$20$\times$4000; this translates to \{$L_x, L_y, L_z$\} 
$\approx$  \{12$\delta_e\times$2$\delta_e\times$500$\delta_e$\} 
for our choices of length units and plasma density. 
The burst duration in light travel length is $L_{GRB} \approx 64\delta_e$.
A relatively flat volume aspect ratio is a compromise to
ensure skin-depth resolution at the lowest possible grid size to give
any 3D structure a chance of growing marginally without spending
excess computing time. Boundary conditions are for the lower (upper)
boundaries: plasma particles are specularly reflected (thermalized)
and outgoing photons are removed. Fields are damped in a layer 
$\Delta L_z \sim 0.1 \% L_z$ on the boundaries to avoid spurious 
wave interference.

Compton scattering is conducted through MC binary 
interaction, using the full Klein-Nishina expression. Photons and 
electrons are selected pairwise at random for scattering according 
to their weights. To achieve detailed balance, scattering is done 
by splitting the particles into a new part carrying the Compton 
scattered fraction of the old part~\citep{bib:Haugboelle2005}, 
and an unscattered remaining part. We may write 
$$w_{(\gamma,e),scatt}=w_{(\gamma,e),old}-w_{(\gamma,e),new}~,$$
where the weights of scattered electrons and photons are 
equal pairwise. The particle scatter fraction per time step then satisfies 
$$w_{(e,\gamma),scatt} \propto w_e w_\gamma \overline{\overline{\sigma_C}} ({\bf p}_e,{\bf p}_\gamma) c \Delta t~,$$
where $\overline{\overline{\sigma_C}} ({\bf p}_e,{\bf p}_\gamma)$ denotes
the microphysical Compton scattering matrix for photons and electrons
in a given cell. For a scattering fraction of $10^{-6}$,
$10^{24}$ \textit{physical} particles are scattered away from a
'mother' particle of size $10^{30}$, for example; this illustrates the
flexible and inherent continuous nature of the \textsc{PhotonPlasma} 
code.

%---------------------------------------------------------------------
\section{Results and Discussion}

\subsection{Wakefield-Plasma Gradient Effects and Transients}\label{subsect:wake-effects} 
\begin{figure}[!t] 
\begin{center} 
\ifnum\epsgz=1
 \epsfig{figure=f2.eps.gz,width=\linewidth} 
\else 
 \epsfig{figure=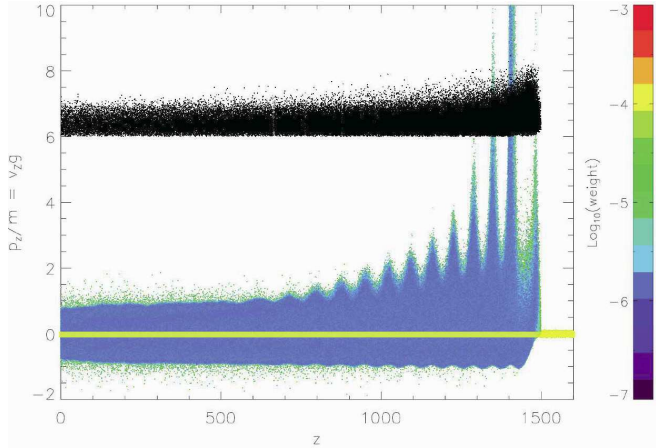,width=\linewidth} 
\fi 
\caption{\textit{Electron Scatter Plot}:  Color designates weight in 
log-scale -- blue particles have weights $w_{blue} \sim 10^{-5}w_0$, 
where $w_0$ is the initial weight. The yellow line coincides with 
zero bulk z-momentum, $p_{z,\gamma}=0$. For clarity photons (black) 
are offset vertically by $p_{z,\gamma}=+6$.} 
\label{fig:logweight1} 
\end{center} 
\end{figure}

\begin{figure}[!t]
\begin{center} 
\ifnum\epsgz=1
 \epsfig{figure=f3.eps.gz,width=\linewidth} 
\else 
 \epsfig{figure=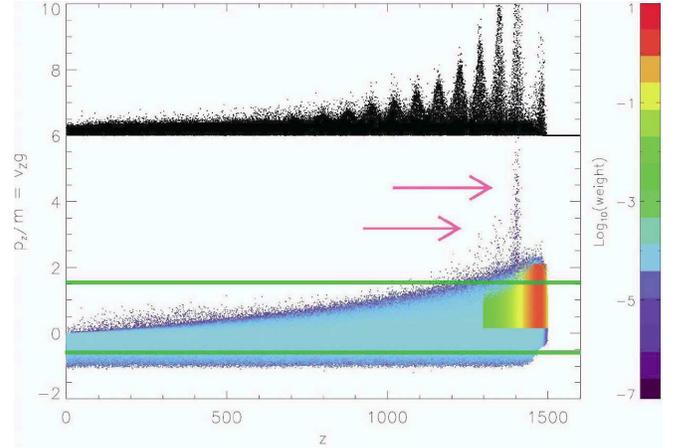,width=\linewidth} 
\fi 
\caption{\textit{Photon Scatter Plot}: same run as 
\fig{fig:logweight1}, but for photons (in color). The initial burst 
(moving right) is visible from $t\sim1500$ to $t\sim1300$. Pink 
arrows: local photon upscattering. Green lines: energy threshold for 
$\gamma+\gamma \rightarrow e^++e^-$ for photons pairwise oppositely 
outside (away from $p_z=0$) these lines. Electrons (black) are offset 
by $p_{z,e}=+6$. Here, $p_{z,\gamma} \equiv [h\nu]_z/m_ec^2$.} 
\label{fig:logweight2} 
\end{center} 
\end{figure}

As the pulse interacts with the quiescent plasma, electrons are strongly
forward Compton scattered. The plasma reacts electrostatically to
restore the displacement by pulling the bulk of the electrons
backwards anti-parallel to the burst propagation direction, while this
is turn back-reacts on the photons. The effects can be seen as the
'dip' from $z=1500$, and to the left, in figures 
\ref{fig:logweight1} and \ref{fig:logweight2}, where scatter plots in
momentum sub-space \{${z},\left|{\bf p}_z\right|$\} are shown for
electrons and photons, respectively. The scatter plots are from runs
with twice the resolution and half the density of those for the main
simulation (Section 2 and fig.\ref{fig:3dfields}). 
The well resolved and longer skin-depth reveals strong 
electrostatic beating in the plasma parallel momentum. These beats 
last of order $\tau_b \sim \delta_e/c \sim 10^{-4}$s in the CBM 
rest frame, and particles caught in the wake reach Lorentz factors 
upward of $\gamma_e \sim 30$. 

Forwardly accelerated plasma -- above the yellow line in 
\fig{fig:logweight1} -- relaxes to $\left|p_{z,e}\right| \sim 
\pm m_ec$ in a pace proportional to the burst duration, in order to 
counter-act the bulk backwards flow (the 'dip') initiated by the photons. The 
plasma undergoes violent accelerations, to maintain 
charge-neutrality. Burst photons keep the electron population 
'inflated' as long as photon free energy and anisotropy is available 
in the plasma; far downstream the 'inflated' non-thermal population 
lasts through the duration of our simulations.

The violent electrostatic acceleration of plasma electrons, due to the
sharp gradient in photon pressure, will produce bremsstrahlung and
synchrotron photons. In subsequent upscattering these photons will
eventually also affect the spectrum~\citep{bib:Barbiellini2006}. More
importantly, millisecond variability could -- in our assumed
environment -- arise directly from the enhanced density variations,
scattering and possibly pair production (leading to higher densities
and more scattering). Such a variability, which is comparable to the
plasma frequency in a medium with density $n_0 \sim 10^{-1}$cm
$^{-3} - 10^0$cm$^{-3}$, could therefore be self-feeding and grow in
strength and duration. We may be observing a weak seed to such 
spike in the spectrum at high energy, marked by the two pink arrows in
\fig{fig:logweight2}. Such photons spikes could facilitate pair production, 
and modify the GRB spectrum as the burst traverses the CBM.

\subsection{Magnetic Field Generation} 
\begin{figure*}%[t] 
\begin{center} 
\ifnum\epsgz=1
 \epsfig{figure=f4.eps.gz,width=\textwidth} 
\else
 \epsfig{figure=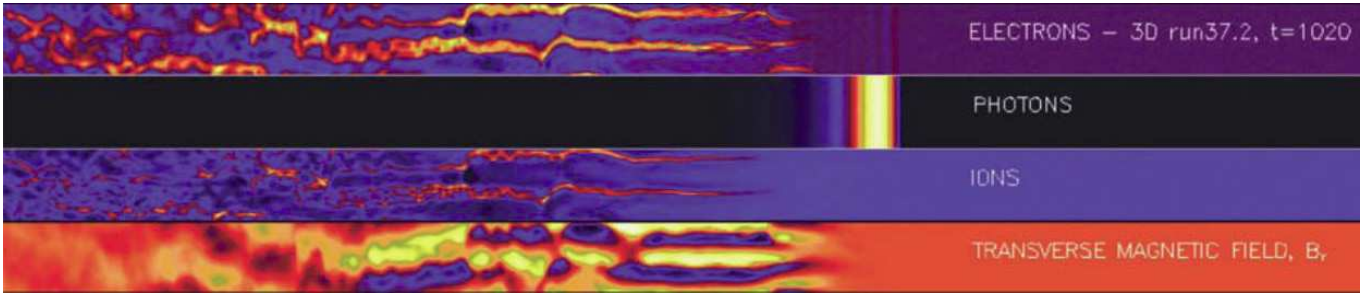,width=\textwidth} 
\fi 
\caption{Current structures for electrons (panel 1), photon pulse 
(panel 2), ions (panel 3) and (transverse, out-of-plane) magnetic 
field structure (panel 4) in our 3D wakefield simulation, at 
simulation time $t=1020\omega_{pe}^{-1}\approx 10^4\Delta t$. The 
GRB photon pulse has light-length $L_c \equiv c/\tau_{GRB} = 200 
\approx 64\delta_e$. The computational volume has dimensions 
\{$L_x, L_y, L_z$\} = \{40, 8, 1600\} or, in electron skin-depths, 
\{$L_x, L_y, L_z$\} $\approx$ 
\{12$\delta_e$, 2$\delta_e$, 500$\delta_e$\}.} 
\label{fig:3dfields}
\end{center} 
\end{figure*}

Most intriguingly, we observe
growth of a strong, large-scale magnetic field in the downstream
wakefield gradient region. The scenario is captured in 
\fig{fig:3dfields}. A cross section of the 3D volume is shown for
electrons, photons, ions and the transverse magnetic field, $B_\perp$
(top-to-bottom), for the 3D run described in Section 2. Barely visible 
in the electron population is the same electrostatic spiky structure 
that shows in \fig{fig:logweight1}, but somewhat weaker due 
comparatively higher density and lower resolution.

From the ion density and electron density panels it is seen that both
species undergo instability to form confluent filaments in the plasma 
(although the inertia difference plays a role). The filamentation build-up 
is initiated by fluctuating EM fields in CBM plasma ahead of the burst. As 
described in Section \ref{subsect:wake-effects}, the plasma forms a 
two-stream anisotropy in reaction to the burst forcing. The fastest growing 
instability in this case is the two-stream filamentation mode -- tantamount 
to the Weibel instability \citep{bib:Weibel1959} -- which produces 
quasi-static magnetic fields. The effect is self-sustaining as long as the 
photon momentum anisotropy is sufficiently high. 

Furthermore, ponderomotive particle forcing,
${\bf F}_p \propto -{q_e^2} {(m\omega)}^{-1} \nabla{\bf E}^2$, is likely to be active 
in the strongly oscillatory wakefield plasma. This drift force is independent 
of the charge sign; both electrons and ions drift in the same manner. Such effects 
has been reported by other authors  \citep{bib:Hoshino2008,bib:Liang2007} to 
being responsible for the acceleration of particles due to forcing by Poynting 
flux pulses. The \textit{coherent} nature of the plasma forcing by these 
authors, contrasts our \textit{stochastic} forcing.

Determining how the magnetic field will decay is a challenge yet to be met; modeling 
a larger plasma volume is necessary to determine the ultimate temporal and spatial
development.  Resorting to lower-dimensional models will not solve the
problem; the instability is inherently 3D in nature and cannot develop realistically 
in lower dimensions \citep[cf.][]{bib:Frederiksen2004}. We also observe in 
\fig{fig:3dfields} that filaments, current structures and field scales are 
limited by the computational volume -- i.e. they are 'boxed-in'. Nonetheless, 
they survive relatively far behind the final tail of the photon wakefield 
(the GRB). Both electrons and ions are responsible for the density filaments, 
the current densities are borne almost entirely by the electrons -- again
-- owing to the inertial difference. We may therefore expect that as
long as the photon anisotropy is "large", the field structures will
survive. 

We proceed to speculate that such magnetized wakefields volumes could provide 
'magnetic walls' against which trailing ejecta could gradually produce a build-
up of shocks and particle acceleration, as the ejecta traverses the
pre-conditioned circumburst medium.

%---------------------------------------------------------------------
\section{Conclusions} We emphasize the importance of bridging the gap
between the (statistical) sub-Debye domain and the macroscopic
Vlasov-/super-Debye domain in astrophysical plasma modeling, opening
the possibility of a detailed kinetic photon description, and allowing
the modeling of processes such as neutron transport and decay,
neutrino streaming and transport, pair production etc. Frequencies for 
wave-particle collisions and for particle-particle 
collisions (e.g. Compton) are comparable in magnitude for 
the physical conditions we have considered in this Letter. In 
effect $$\nu_{\rm coll} \propto \omega_{pe} = c^{-1} 
\delta_e ~~,~~ \nu_{\rm bin} \propto n c \sigma_T ~~
\Rightarrow  ~~ \nu_{\rm coll} \sim \nu_{\rm bin}~,$$ where $n$ is the
photon number density, $\sigma_T$ the Thompson cross section, $l$ the
burst light duration. Subscripts "coll" and "bin" denote "collective"
(wave-particle), and "binary" (particle-particle) - respectively.

Our main findings from these first simplified GRB-CBM wakefield
interaction simulations, which we have conducted with the new 
\textsc{PhotonPlasma} hybrid scheme described briefly in Section 2, 
are: 

\noindent 1. Strong photon density gradients plays an important role
in the dynamics of the CBM while being traversed by a GRB. We
therefore lend support from our studies to suggestions set forth by
\citep{bib:Barbiellini2006} to study in more detail these
stochastically induced wakefield processes.

\noindent 2. The GRB spectrum is locally affected by the same gradient
density induced effects, and this leads to local high energy
fluctuations in the GRB spectrum and light curve on
sub-millisecond-to-millisecond time scales.

\noindent 3. A transverse magnetic field, quasi-static in the CBM frame, 
is generated in the wakefield of a GRB traversing a quiescent 
plasma. The field scales are likely limited by our numerical capabilities 
to follow the growth.

It has been argued theoretically~\citep{bib:Beloborodov2002,
bib:Madau2000} that the CBM ahead of the trailing
forward shock could be loaded with a relatively high density pair
plasma component. From this work, and from observing 
\fig{fig:logweight2}, we see that accounting for pair production 
($\gamma + \gamma \leftrightarrow e^+ + e^-$) is imperative. In 
future work we shall add such dynamics, to test in particular
whether density fluctuations could grow and eventually lead to 
externally produced shocks as a consequence of $\gamma$-ray 
preconditioning of the circumburst medium.

%---------------------------------------------------------------------
\acknowledgements Collaboration with Troels Haugb{\o}lle, Christian
Hededal, and {\AA}ke Nordlund on the development of the
\textsc{PhotonPlasma} code  is acknowledged. Computer time was
provided by the Danish Center for Scientific Computing (DCSC).

%---------------------------------------------------------------------

%---------------------------------------------------------------------
\end{document}